\documentclass[letter]{aa} 
\usepackage{graphicx}
\usepackage{txfonts}

\begin{document}
   \title{
The strong magnetic field of the large-amplitude $\beta$\,Cephei pulsator V1449\,Aql\thanks{Based
on observations obtained at the 2.56-m Nordic Optical Telescope on La Palma and ESO Prgs.\ 077.D-0311 and 178.D-0361.}}
\titlerunning{The magnetic field of V1449\,Aql}

   \author{
S.~Hubrig\inst{1}
\and I.~Ilyin\inst{1}
\and M.~Briquet\inst{2,3}
\and M.~Sch\"oller\inst{4}
\and J.~F.~Gonz\'alez\inst{5}
\and N.~Nu\~nez\inst{5}
\and P. De Cat\inst{6}
\and T. Morel\inst{7}
}

\authorrunning{Hubrig et al.}

\institute{Leibniz-Institut f\"ur Astrophysik Potsdam (AIP), An der Sternwarte 16, 14482 Potsdam, Germany\\
              \email{shubrig@aip.de}
\and Instituut voor Sterrenkunde, Katholieke Universiteit Leuven, Celestijnenlaan 200 D, 3001~Leuven, Belgium
\and LESIA, Observatoire de Paris, CNRS, UPMC, Universit\'e Paris-Diderot, 92195 Meudon, France 
\and European Southern Observatory, Karl-Schwarzschild-Str.\ 2, 85748 Garching bei M\"unchen, Germany        
\and Instituto de Ciencias Astronomicas, de la Tierra, y del Espacio (ICATE), 5400 San Juan, Argentina 
\and Koninklijke Sterrenwacht van Belgi\"e, Ringlaan 3, 1180 Brussel, Belgium
\and Institut d'Astrophysique et de G\'eophysique, Universit\'e de Li\`ege, All\'ee du 6 Ao\^ut, B\^at. B5c, 4000~Li\`ege,
 Belgium
}
\date{Received; accepted}

 
  \abstract
   {}
   {
Only for very few $\beta$~Cephei stars has the behaviour of the magnetic field been studied over the rotation cycle.
During the past two years we have obtained multi-epoch polarimetric spectra of the $\beta$~Cephei
star V1449\,Aql with SOFIN at the Nordic Optical Telescope
to search for a rotation period and to constrain the geometry of the magnetic field.
}
   {
The mean longitudinal magnetic field is measured at 13 different epochs.
The new measurements, together with the previous FORS\,1 
measurements, have been used  for the frequency analysis and the characterization of the magnetic field.
}
   {
V1449\,Aql so far possesses the strongest longitudinal magnetic field of up to 700\,G among
the $\beta$\,Cephei stars.
The resulting periodogram 
displays three dominant peaks with the highest peak at $f=0.0720$\,d$^{-1}$ 
corresponding to a period $P=13\fd893$.
The magnetic field geometry can likely be described 
by a centred dipole with a polar magnetic field strength $B_{\rm d}$ around 3\,kG and an 
inclination angle $\beta$ of the magnetic axis to the rotation axis of 76$\pm$4$^{\circ}$. 
As of today, the strongest longitudinal magnetic fields are detected in the $\beta$\,Cephei stars
V1449\,Aql and $\xi^1$\,CMa with large radial velocity amplitudes.
Their peak-to-peak amplitudes reach $\sim$90\,km\,s$^{-1}$ and $\sim$33\,km\,s$^{-1}$,
respectively.
Concluding, we briefly discuss the position of the currently known eight magnetic $\beta$\,Cephei
and candidate $\beta$\,Cephei stars in the Hertzsprung-Russell (H-R) diagram.
}
   {}


   \maketitle
%

\section{Introduction}

Before the CoRoT mission, the B1.5~II-III star V1449\,Aql (=HD\,180642)
was known as a $\beta\,$Cephei pulsator with a 
high-amplitude radial mode of frequency 5.487\,d$^{-1}$
(Waelkens et al.\ \cite{Waelkens1998}).
White light photometry from space
revealed the rich frequency spectrum of this target (Degroote et al.\ \cite{Degroote2009}) with the first detection of 
solar-like oscillations in a massive star (Belkacem et al.\ \cite{Belkacem2009}).
Additional ground-based, multi-colour 
photometry and high-resolution spectroscopy were used to put constraints on the atmospheric parameters
and the chemical abundances, and to 
identify three modes of low amplitudes (Briquet et al.\ \cite{Briquet2009}). 
Notably, the dominant mode does not behave sinusoidally, and the first velocity moment $\langle v^1\rangle$ has a peak-to-peak 
amplitude of about 90 km/s. This is the third largest peak-to-peak radial--velocity amplitude measured among  $\beta~$Cephei stars,
after BW\,Vul
(e.g.\ Crowe \& Gillet \cite{CroweGillet1989})
and $\sigma$~Sco (Mathias et al.\ \cite{Mathias1991}).
A detailed NLTE abundance analysis showed a nitrogen excess of $\sim$0.3\,dex with respect to the average nitrogen abundance in B-type stars in 
the solar neighbourhood. Previous abundance studies of hot stars indicate that the appearance of such an excess may be 
linked to the presence of a magnetic field (e.g., Morel et al.\ \cite{Morel2006,Morel2008}; Przybilla \& Nieva \ \cite{PrzybillaNieva2010}).

The first spectropolarimetric observations of V1449\,Aql were obtained  with the multi-mode instrument FORS\,1  at the VLT
on two consecutive nights in 2007 and revealed a change in polarity from one night to the next and the presence
of a weak magnetic field 
on the second night of observations (Hubrig et al.\ \cite{Hubrig2009}).
In the present work we discuss 13 new spectropolarimetric observations
acquired during the last two years at the 2.56\,m Nordic Optical Telescope using the SOFIN echelle spectrograph.
The new longitudinal magnetic field measurements are used to determine the stellar rotation period and to put 
constraints on the magnetic field geometry. 

\section{Magnetic field measurements and period determination}



Multi-epoch series of polarimetric spectra of V1449\,Aql with S/N $\ge$200 were obtained 
with the low-resolution camera ($R=\lambda/\Delta\lambda\approx30000$)
of the high-resolution echelle spectrograph SOFIN (Tuominen et al.\ \cite{Tuominen1999})
mounted at the Cassegrain focus of the Nordic Optical Telescope (NOT)
in 2009 August--September and in 2010 July.
We used a 2K Loral CCD detector to register 40 echelle orders partially covering
the range from 3500 to 10\,000\,\AA{} 
with a length of the spectral orders of about 140\,\AA{} at 5500\,\AA{}.
Two exposures with the quarter-wave plate angles separated 
by $90^\circ$ are necessary to derive circularly polarized spectra.
The spectra are reduced with the 4A software package (Ilyin \cite{Ilyin2000}). 
Wavelength shifts between the right- and left-hand side circularly polarized spectra were
interpreted in terms of a longitudinal magnetic field $\left<B_{\rm z}\right>$,
using the moment technique described by Mathys (\cite{Mathys1994}).


\begin{table}
\centering
\caption{
Magnetic field measurements of V1449\,Aql with SOFIN.
}
\label{tab:log_meas}
\begin{tabular}{ccr@{$\pm$}l|ccr@{$\pm$}l}
\hline
\multicolumn{1}{c}{JD --} &
\multicolumn{1}{c}{Phase} &
\multicolumn{2}{c}{$\left<B_{\rm z}\right>$} &
\multicolumn{1}{c}{JD --} &
\multicolumn{1}{c}{Phase} &
\multicolumn{2}{c}{$\left<B_{\rm z}\right>$} \\
\multicolumn{1}{c}{2450000} &
&
\multicolumn{2}{c}{[G]} &
\multicolumn{1}{c}{2450000} &
&
\multicolumn{2}{c}{[G]} \\
\hline
\hline
4343.659 & 0.790 & $-$55  & 33 & 5397.567 & 0.648 & $-$430 & 89 \\
4344.584 & 0.856 &   166  & 41 & 5398.530 & 0.717 & $-$255 & 211 \\
5071.396 & 0.171 & $-$21 & 95 & 5400.574 & 0.864 &   168 & 168 \\
5073.391 & 0.314 & $-$380 & 159 & 5402.577 & 0.008 &   421 & 115 \\
5076.389 & 0.530 & $-$647 & 153 & 5405.646 & 0.229 & $-$204 &131 \\
5078.429 & 0.677 & $-$326 & 171 & 5407.559 & 0.367 & $-$416 & 133 \\
5080.392 & 0.818 & $-$115 & 102 & 5409.648 & 0.517 & $-$663 & 427 \\
5395.564 & 0.504 & $-$729 & 232 \\
\hline
\end{tabular}
\tablefoot{
In the first two lines we list the earlier FORS\,1 measurements 
published by Hubrig et al.\ (\cite{Hubrig2009}).
Phases are calculated according to the ephemeris of 
${\rm JD} = 2455305.21 + 13.893~{\rm E}$.
All quoted errors are 1$\sigma$ uncertainties.
}
\end{table}

The magnetic field measurements with SOFIN and FORS\,1 are presented in Table~\ref{tab:log_meas}.
In the first column we list the heliocentric 
Julian dates for the middle of the spectropolarimetric observations. The phases of the measurements of 
the magnetic field are shown in Col.~2, and in Col.~3 we present the mean longitudinal magnetic 
field $\left<B_{\rm z}\right>$.
An example of Stokes~$I$ and Stokes~$V$ profiles of \ion{N}{II} in the 
wavelength region $\lambda\lambda$ 4775--4800 is presented in Fig.~\ref{fig:new}.

\begin{figure}
\centering
\includegraphics[width=0.40\textwidth]{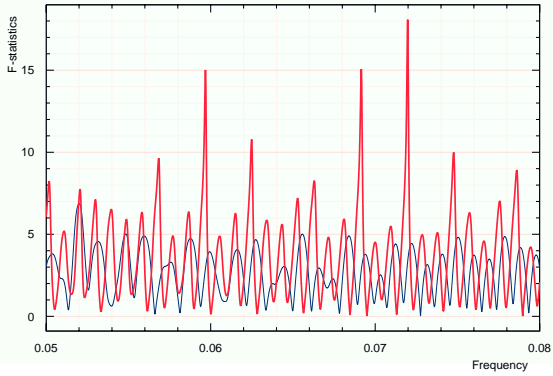}
\caption{
The periodogram of V1449\,Aql built from the longitudinal magnetic field measurements
with the most prominent frequency at $f=0.0720$\,d$^{-1}$, corresponding to
a period of 13.893\,days. The lower thin line presents the window function. }
\label{fig:period}
\end{figure}

\begin{figure}
\centering
\includegraphics[width=0.35\textwidth]{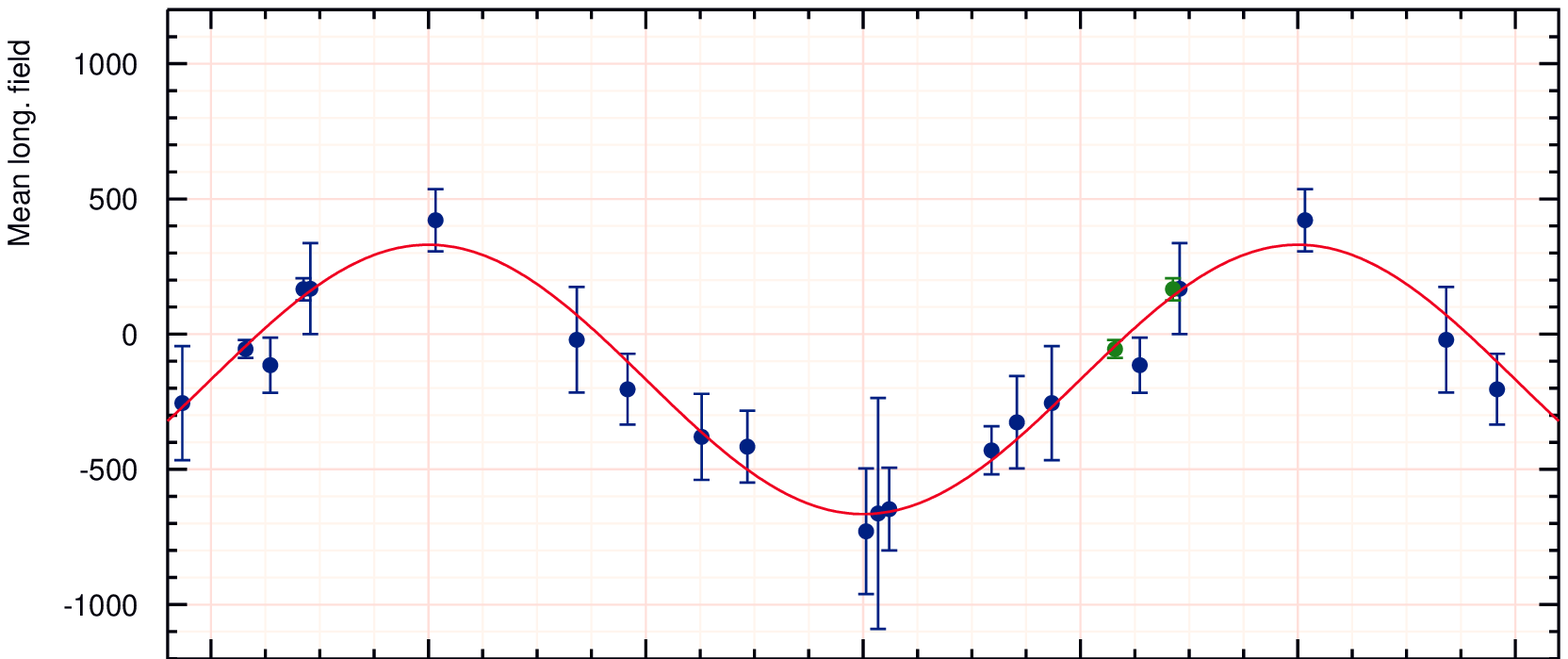}
\includegraphics[width=0.35\textwidth]{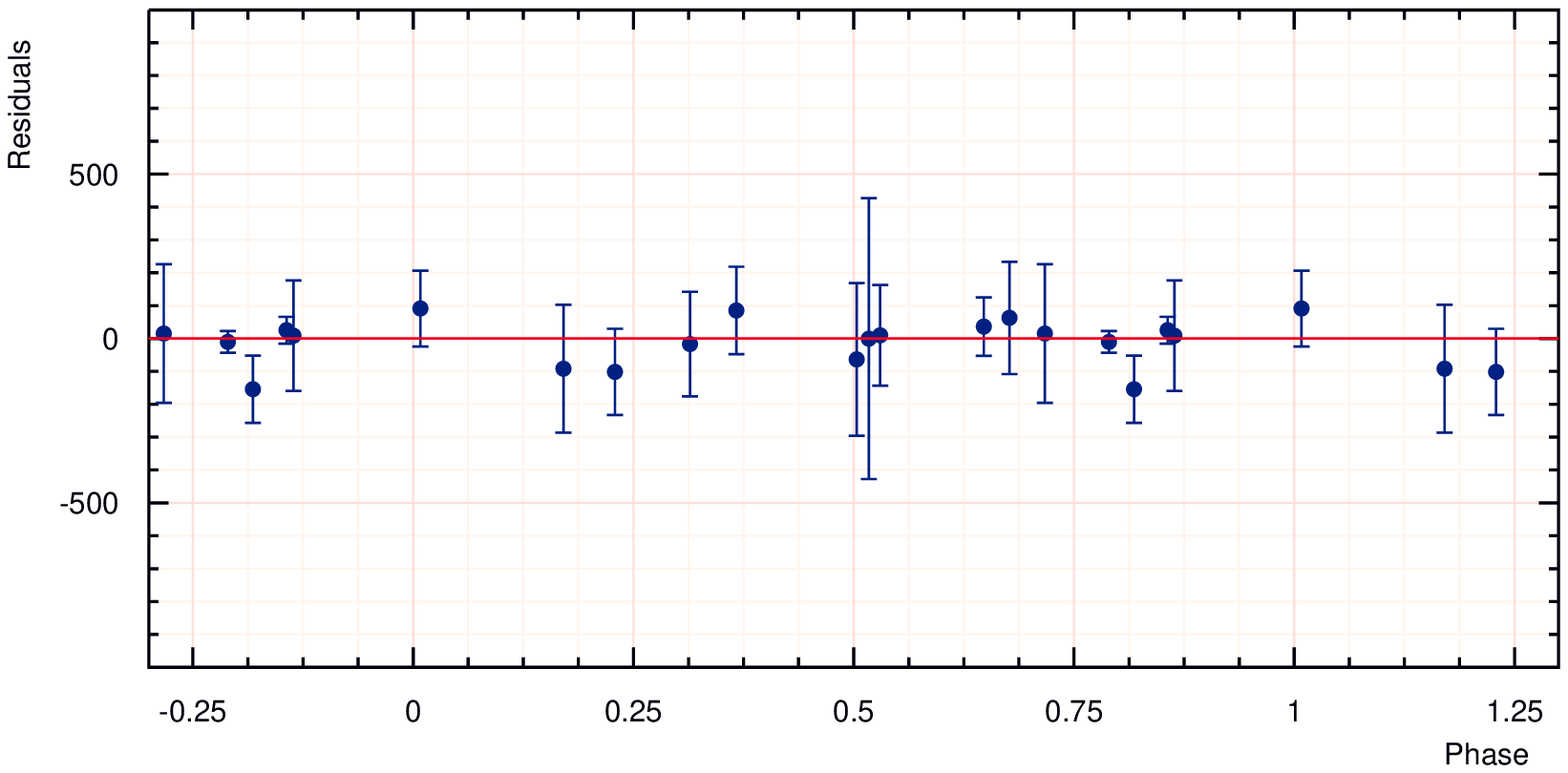}
\caption{
Phase diagram with the best sinusoidal fit for the longitudinal magnetic field measurements
with a reduced $\chi^2=0.46$.
The residuals (Observed -- Calculated) are shown in the lower panel.
The two FORS\,1 measurements appear in green. 
}
\label{fig:Bz}
\end{figure}

The frequency analysis  was performed on the 13 SOFIN longitudinal magnetic field 
measurements and two previous FORS\,1 measurements using a non-linear least-squares fit of the multiple harmonics 
utilising the Levenberg-Marquardt 
method (Press et al.\ \cite{Press92}) with an optional possibility of pre-whitening the trial harmonics.  
To detect the most probable period, we calculated the frequency spectrum and for each trial frequency we performed a 
statistical F-test of the null hypothesis for the absence of periodicity (Seber \cite{Seber77}).
The resulting periodogram 
displays three dominant peaks with the highest peak at $f=0.0720$\,d$^{-1}$ 
corresponding to a period $P=13\fd893$.
In the framework of the rigid rotator model usually assumed for magnetic stars
on the upper main-sequence,
the period of the magnetic field variation corresponds to the stellar rotation period.
In Fig.~\ref{fig:period} we show that two lower peaks appear at $f=0.069162$\,d$^{-1}$ and $f=0.059689$\,d$^{-1}$
with equivalent periods of $P=14\fd459$ and $P=16\fd753$, respectively. Since two measurements in our data set were obtained 
with the low-resolution FORS\,1 spectrograph, we carried out a few tests to evaluate the significance of these 
measurements for the period 
determination. We find that omitting the FORS\,1 measurements
from the period search leads to the dominance of a frequency peak corresponding
to a period of $14\fd496$. 
On the other hand, the period search without weighting tied to the measurement accuracies 
leads to a much stronger dominance of the frequency peak $f=0.0720$\,d$^{-1}$ 
with the equivalent period of $P=13\fd893$. 
The referee of this paper informed us that his implementation of a CLEAN analysis leads to 
a detection of the same periods, but favors the period of $P=14\fd459$.
These results show that the periodicity determination may depend on the technique used and that 
the relevance of the frequencies $f=0.069162$\,d$^{-1}$ 
and $f=0.059689$\,d$^{-1}$ is difficult to ascertain at the present stage without additional magnetic data.
The derived ephemeris for the detected period corresponding to the highest frequency peak is 
\begin{equation}
\left<B_{\rm z}\right>^{\rm pos. extr.} = {\rm HJD} 2455305.21 \pm 0.16 + 13.893 \pm 0.003E.
\end{equation}
The phase diagram of the magnetic field measurements
for the determined period is presented in Fig.~\ref{fig:Bz}.

\begin{figure}
\centering
\includegraphics[width=0.35\textwidth]{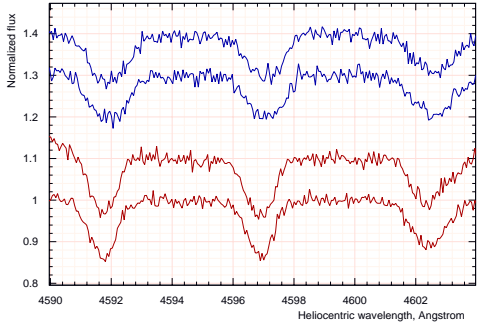}
\caption{
Variability of the output spectra in two SOFIN subexposures taken with
the quarter-wave plate angles separated by $90^\circ$
taken around HJD\,2455398.530.
The lower two spectra, $(I+V)_0$ and 
$(I-V)_0$, correspond to the first subexposure, while the upper spectra, $(I-V)_{90}$ and $(I+V)_{90}$, correspond to the 
second subexposure.
The strong effect of pulsations on the line profile shapes and the line positions is clearly visible 
between the spectra of the first subexposure with a duration of 20\,min and the spectra of the second subexposure 
with the same duration. 
}
\label{fig:meas}
\end{figure}

The measurements of the longitudinal magnetic field in V1449\,Aql require,
however, special care in treating the data, as  
the timescale of a complete single observation of the length of $\sim$40 min with SOFIN,
including two subexposures, 
is comparable to the timescale of the pulsation variability. 
As we mention above, two subexposures with the quarter-wave plate angles separated 
by $90^\circ$ are needed to derive circularly polarized spectra. To remove the instrumental effects of small misalignments, 
differences in transmission, etc., the complete observation of a star usually consists of at least two 
successive subexposures with the waveplate rotated by $90^\circ$ to exchange the positions of the two spectra on the CCD.
The final Stokes $I$ and $V$ spectra are then calculated from the polarized spectra by taking the respective averages.
To obtain Stokes $I$, the two average corresponding spectra, in our case the average of  
$(I+V)_0$ and  $(I+V)_{90}$ are added with the average of $(I-V)_0$ and $(I-V)_{90}$,  while Stokes $V$ is 
calculated from the difference of the two spectra, the average of $(I+V)_0$ and  $(I+V)_{90}$ and of $(I-V)_0$ and $(I-V)_{90}$.
However, owing to very strong changes in the line profile positions and shapes
in the strongly pulsating V1449\,Aql, the method of 
using respective average spectra would lead to erroneous wavelength shifts and thus to wrong values of the longitudinal magnetic field. 
The behaviour of all four spectra in both subexposures
taken around HJD\,2455398.530 is illustrated in Fig.~\ref{fig:meas}. 
To remedy this situation, we decided to carry out the measurements of the line shifts between the spectra $(I+V)_0$ and 
$(I-V)_0$ in the first subexposure and the spectra $(I-V)_{90}$ and  $(I+V)_{90}$  in the second subexposure separately. The final 
longitudinal magnetic field value was then calculated as an average of the measurements for each subexposure.

\begin{figure}
\centering
\includegraphics[angle=270,width=0.40\textwidth]{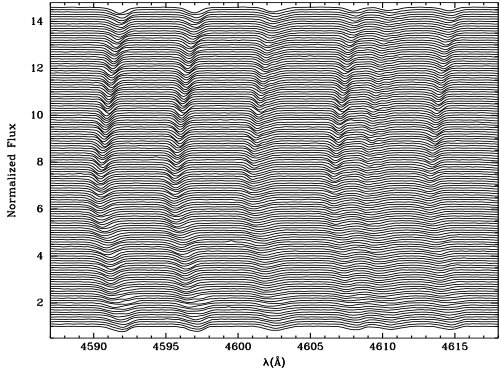}
\caption{
Time series of FEROS spectra showing pulsational line profile variability in the spectral region 4590--4615\,\AA{}.
The pulsation phase zero is at the bottom.
}
\label{fig:feros}
\end{figure}

\begin{figure}
\centering
\includegraphics[height=0.45\textwidth,angle=270]{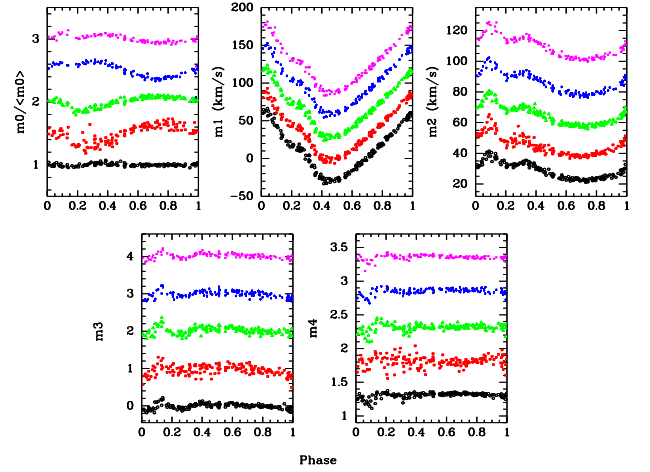}
\caption{
Velocity moments of line profiles: equivalent width (m0), radial velocity (m1),
line width (m2), asymmetry (m3), and kurtosis (m4). In each panel five curves
are plotted corresponding to five spectral lines of various elements:
\ion{He}{I} 4713.1, \ion{C}{II} 5133.1, \ion{N}{II} 5667.6, 
\ion{O}{II} 4592.0, and \ion{Si}{III} 4553.6, shown from bottom to top.
}
\label{fig:moment}
\end{figure}

The line profiles undergo very significant 
pulsational variability, and any rotational modulation of the observed profiles is completely masked by the pulsational
modulation.  In Fig.~\ref{fig:krivky} we demonstrate the behaviour of the \ion{O}{ii} and \ion{He}{i}
line profiles in SOFIN spectra in the spectral region 4695--4715\,\AA{} over the rotation cycle. 
The inspection of 224 FEROS spectra used in the study of Briquet et al.\ (\cite{Briquet2009}) reveals that all 
lines behave in a rather similar way. In Fig.~\ref{fig:feros} we present a clear pulsation pattern in the spectral
region 4590--4615\,\AA{}. The moment variations for five lines belonging to different elements, namely
\ion{He}{I} 4713.1, \ion{C}{II} 5133.1, \ion{N}{II} 5667.6, \ion{O}{II} 4592.0, and \ion{Si}{III} 4553.6,
shown in Fig.~\ref{fig:moment}, however, exhibit 
slight differences between the elements. The equivalent widths of \ion{O}{II} and \ion{Si}{III} are larger at the phase of
minimum radius, while \ion{C}{II} and \ion{N}{II} are larger at the maximum radius. 
The equivalent width of \ion{He}{I} is almost constant.  At minimum radius the elements N, C, and O show
line profiles narrower than those of Si and He. Also the amplitude of the radial velocity curve of the \ion{He}{I} line 
is about 4\,km\,s$^{-1}$ greater than for the \ion{C}{II} line and 2\,km\,s$^{-1}$ more than for the \ion{N}{II} line.

\begin{table}
\centering
\caption{
Magnetic field model for V1449\,Aql.
}
\label{tab:dipvals}
\begin{tabular}{cc|r@{$\pm$}lr@{$\pm$}lr@{$\pm$}lr@{$\pm$}l}
\hline
\hline
$\overline{\left< B_{\rm z}\right>}$ & [G]           & $-$167.5& 27.3 \\ 
$A_{\left< B_{\rm z}\right>}$        & [G]           & 498.1 & 37.1 \\
$P$                                  & [d]           & 13.893 & 0.003 \\
$R$                                  & [$R_{\odot}$] & 11.2 & 1.4 \\
$v_{\rm eq}$                         & [km s$^{-1}$] & 41 & 5 \\
$v$\,sin\,$i$                        & [km s$^{-1}$] & 24 & 2 \\
$i$                                  & [$^{\circ}$]  & 36 &  6 \\
$\beta$                              & [$^{\circ}$]  & 76 &  4 \\
$B_{\rm d}$                         & [G]           & 3100 & 300 \\
\hline
\end{tabular}
\end{table}

The longitudinal field of V1449\,Aql can be adequately represented
by a single-wave variation curve during the stellar rotation cycle (see Fig.~\ref{fig:Bz}),
indicating a dominant dipolar contribution to the magnetic field 
topology. 
Assuming that V1449\,Aql is an oblique dipole rotator, 
the magnetic dipole axis tilt $\beta$ is constrained by 
$\left< B_{\rm z}\right>^{\rm max}/\left< B_{\rm z}\right>^{\rm min} = \cos(i+\beta)/\cos(i-\beta)$.
 %
 %
In Table~\ref{tab:dipvals} we show 
the mean value $\overline{\left< B_{\rm z}\right>}$ and the semi-amplitude of the field variation $A_{\left< B_{\rm z}\right>}$ for V1449\,Aql in rows~1 and 2.
Values for $v_{\rm t}=44$\,km\,s$^{-1}$, which is the total line broadening,
and $v_{\rm eq}=38\pm15$\,km\,s$^{-1}$ were published by Briquet et al.\ (\cite{Briquet2009}). 
The stellar radius,  $R=11.2\pm1.4$\,$R_\odot$, was calculated using main-sequence evolutionary CL\'ES
models (Scuflaire et al.\ \cite{Scu2008}).
Obviously, the accuracy of the determination of $v_{\rm eq}$ 
from the best-fit-solutions of the spectroscopic 
mode identification described in the work of Briquet et al.\ (\cite{Briquet2009}) is rather low. 
However, the equatorial rotation velocity can also be calculated using the relation
$v_{\rm eq}=50.6$\,$R/P$, where $R=11.2\pm1.4$\,$R_\odot$
is the stellar radius in solar units and $P=13.893$\,d the period in days. 
Using this formula, we obtain $v_{\rm eq}=41\pm5$\,km\,s$^{-1}$.
On the other hand, Lefever (private communication) obtained $v$\,sin\,$i=24$\,km\,s$^{-1}$,
using the different procedure described in Lefever et al.\ (\cite{Lefever2010}), which takes macroturbulent broadening into account. 
Our own estimate of the  $v$\,sin\,$i$ value, $v$\,sin\,$i=23.5$\,km\,s$^{-1}$ using the technique of D\'iaz 
et al.\ (\cite{Diaz2010}) fully agrees with that of Lefever.  
The combination of $v$\,sin\,$i=24$\,km\,s$^{-1}$ with a typical estimation error of 2\,km\,s$^{-1}$
and $v_{\rm eq}=41\pm5$\,km\,s$^{-1}$ leads to the inclination angle $i=36\pm6^{\circ}$.
In the last two rows, we list the parameters of the magnetic field dipole model.

\section{Discussion}

Using FORS\,1/2 and SOFIN longitudinal magnetic field
measurements collected in our recent studies,
we were able to detect a weak mean longitudinal magnetic field 
of a few hundred Gauss in four $\beta$\,Cephei stars,
$\delta$\,Cet, $\xi^1$\,CMa, 15\,CMa, and V1449\,Aql,
and in two candidate $\beta$\,Cephei stars, $\alpha$~Pyx and $\epsilon$~Lup
(Hubrig et al.\ \cite{Hubrig2006,Hubrig2009,Hubrig2011}). 
Before we started our systematic search for magnetic fields in pulsating B-type stars,
a weak magnetic field was detected in two other $\beta$\,Cephei stars,
in the prototype of the class, $\beta$\,Cep itself, 
by Henrichs et al.\ (\cite{Henrichs2000}) and in V2052\,Oph by Neiner et al.\ (\cite{Neiner2003}).
The strongest longitudinal magnetic fields were detected in the two $\beta$\,Cephei stars
with the large radial velocity amplitudes, namely in V1449\,Aql and $\xi^1$\,CMa,
for which peak-to-peak amplitudes reach $\sim$90\,km\,s$^{-1}$ and $\sim$33\,km\,s$^{-1}$,
respectively.
The rather strong magnetic field 
in the atmosphere of these two stars and the rather low  $v$\,sin\,$i$ values,
certainly make these stars the most suitable targets for studying  the impact of magnetic fields on stellar rotation, 
pulsations, element diffusion, and frequency patterns. 

In the frequency analysis of the CoRoT data obtained for V1449\,Aql (Degroote et al.\ \cite{Degroote2009}), 
the frequency $f=0.068790\pm0.000393$\,d$^{-1}$ is, within the errors, close to the frequency $f=0.069162$\,d$^{-1}$
detected in our study of 
the periodicity using magnetic field measurements. Additional magnetic field measurements will be worthwhile
for proving whether this photometric frequency corresponds to the rotation/magnetic period of this star.  

\begin{figure}
\centering
\includegraphics[height=0.35\textwidth,angle=270]{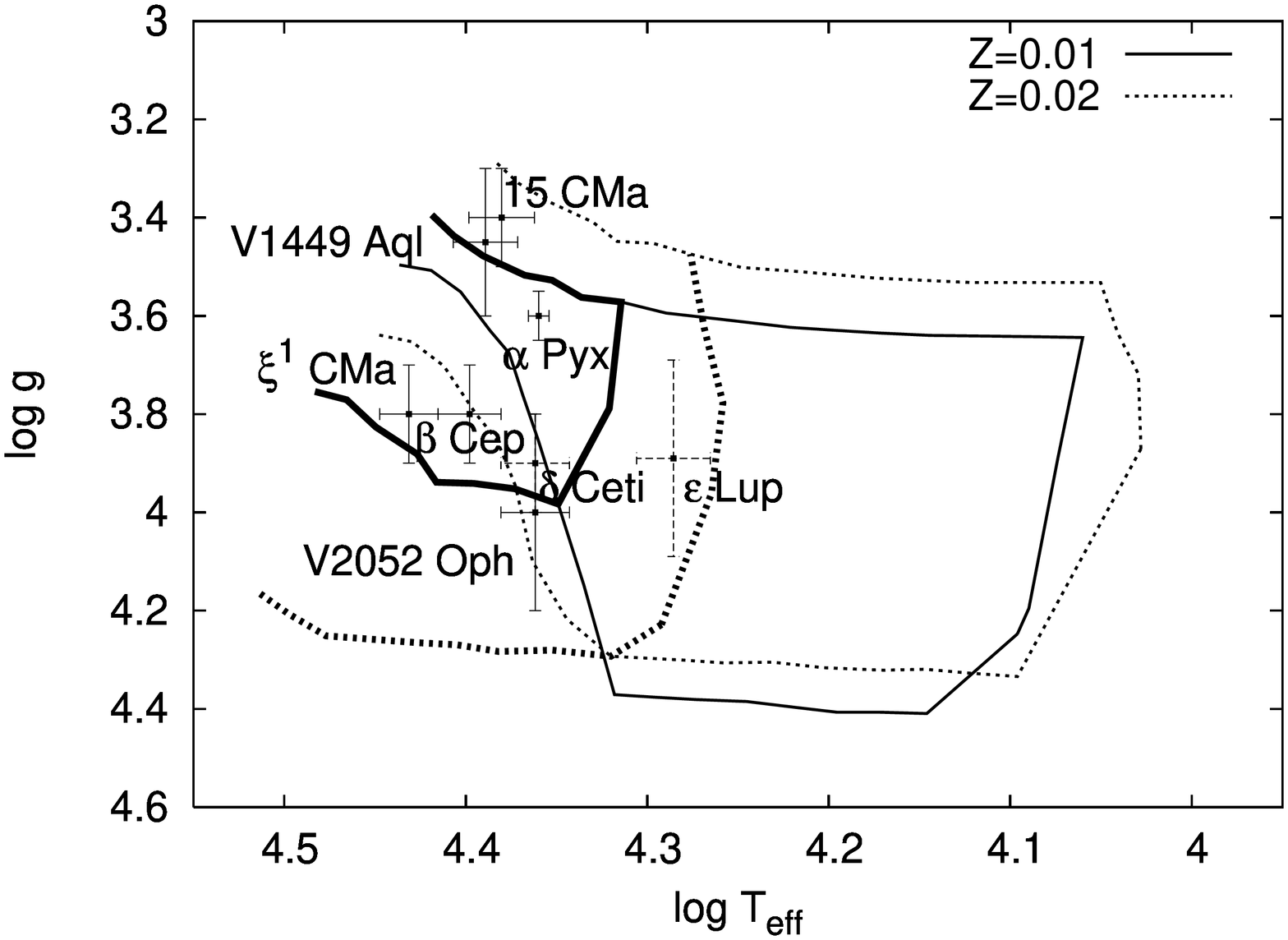}
\includegraphics[height=0.35\textwidth,angle=270]{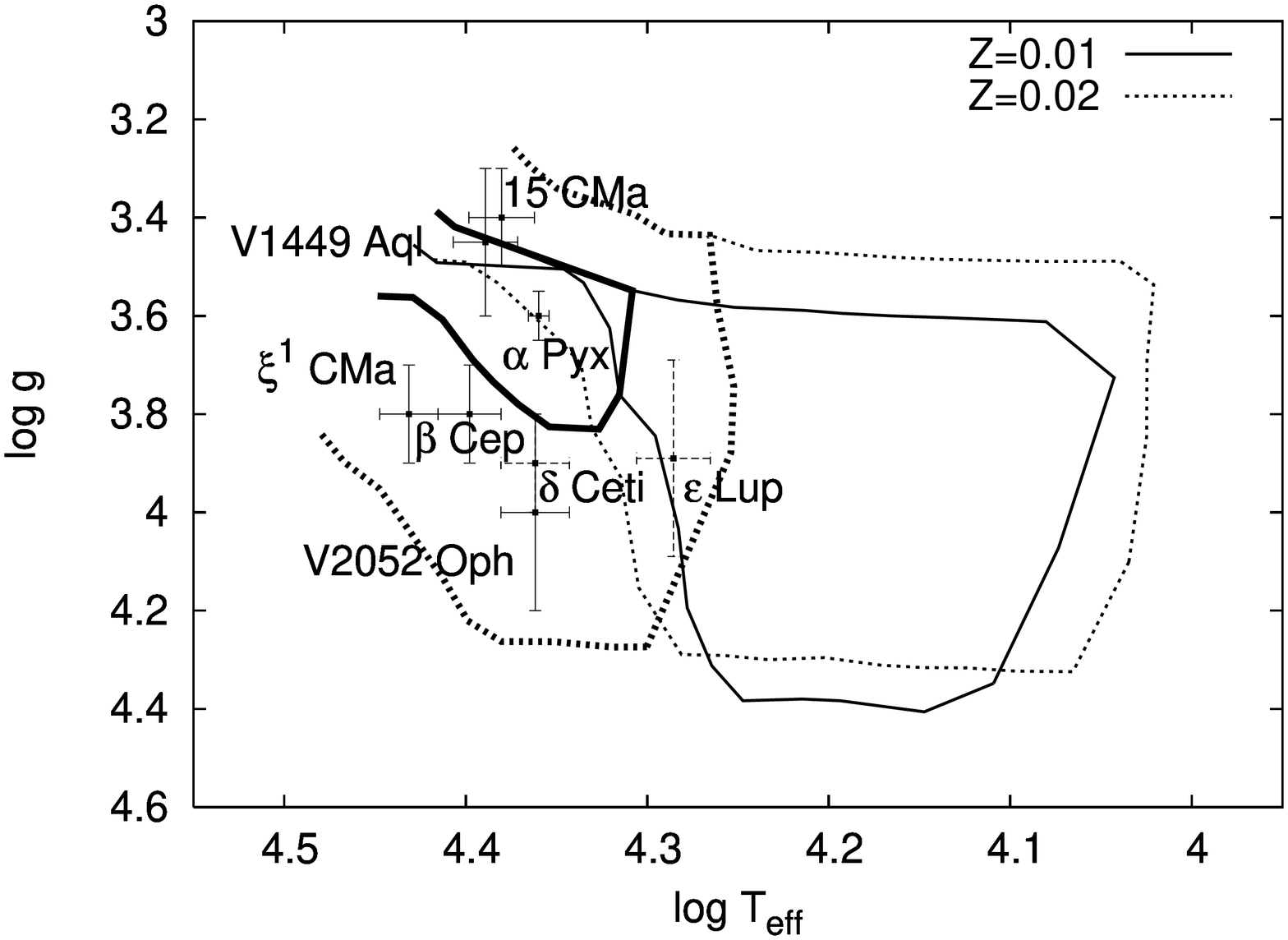}
\caption{The position of the $\beta$\,Cephei and candidate $\beta$\,Cephei stars in the H-R diagram.
Upper panel: The boundaries of the theoretical instability strips are calculated 
using the OP opacities.
Lower panel: Boundaries are calculated using OPAL opacities. Full lines correspond to strips for
metallicity $Z=0.01$ and dotted lines to strips with metallicity $Z=0.02$. The thick and thin lines 
correspond to the boundaries of the $\beta$~Cephei and SPB instability regions, respectively. 
}
\label{fig:m6149}
\end{figure}


Our studies indicate that dipole models provide a satisfactory fit to the magnetic data and
among the  presently known magnetic $\beta$\,Cephei stars, 
$\xi^1$\,CMa and V1449\,Aql, possess the largest magnetic fields, with a dipole strength of several kG. 
The position of the currently known eight magnetic $\beta$\,Cephei and candidate $\beta$\,Cephei stars in 
the log\,T$_{\rm eff}$--log\,$g$ diagram is presented in Fig.~\ref{fig:m6149} 
together with the boundaries of the theoretical instability strips calculated for different metallicities ($Z=0.01$ and $Z=0.02$)
and using the OP opacities (http://cdsweb.u-strasbg.fr/topbase/op.html, see also Miglio et al.\ \cite{Miglio2007}) and  
OPAL opacities 
(http://opalopacity.llnl.gov/opal.html). The stellar fundamental parameters and their literature sources are compiled
in Table~\ref{tab:fundamental}.
We give in the same table the information on the dominant mode and the radial velocity
amplitude, if known. 
The extent and the position of the instability strip are highly dependent 
on the opacities used and on the metallicity adopted. Recent studies indicate, however, a lower metallicity than
$Z=0.02$ for early B-type stars in the solar neighbourhood with $Z=0.014$
(e.g.\ Przybilla et al.\ \cite{Przybilla2008}).
Although the sample of magnetic pulsating stars is still very small, and it is difficult to determine 
any trend in the distribution of the stars in the H-R diagram, it might seem that magnetic stars are clustered 
rather close to the instability strip boundaries
(full lines in left panel of Fig.~\ref{fig:m6149}).
Certainly, additional future magnetic field studies of a larger sample of 
$\beta$\,Cephei stars are needed to study the indicated trend.

\begin{acknowledgements}
M.~B. acknowledges the Fund for 
Scientific Research, Flanders, for a grant for a long stay abroad. 
T.~M. acknowledges financial support from Belspo for contract PRODEX-GAIA DPAC.
We thank the anonymous referee for useful comments and suggestions.
\end{acknowledgements}

\appendix

\section{Additional material}

\begin{table*}
\centering
\caption{
Fundamental parameters of $\beta$\,Cephei and candidate $\beta$\,Cephei stars with detected magnetic fields.
}
\label{tab:fundamental}
\begin{tabular}{lr@{$\pm$}lr@{$\pm$}lcccc}
\hline
\hline
\multicolumn{1}{c}{Object} &
\multicolumn{2}{c}{$T_{\rm eff}$} &
\multicolumn{2}{c}{log\,$g$} &
\multicolumn{1}{c}{Reference} &
\multicolumn{1}{c}{dominant mode} &
\multicolumn{1}{c}{RV amplitude} &
\multicolumn{1}{c}{Reference} \\
\multicolumn{1}{c}{name} &
\multicolumn{2}{c}{[K]} &
\multicolumn{2}{c}{} &
\multicolumn{1}{c}{} &
\multicolumn{1}{c}{} &
\multicolumn{1}{c}{[km s$^{-1}$]} &
\multicolumn{1}{c}{} \\
\hline
$\delta$\,Cet   & 23000 & 1000 & 3.90  & 0.10 &  1 & $f=6.2026$\,d$^{-1}$  & 7.44 & 5 \\
$\xi^1$~CMa     & 27000 & 1000 & 3.80  & 0.10 &  1 & $f=4.77153$\,d$^{-1}$  & 16.4 & 6 \\
15\,CMa         & 24000 & 1000 & 3.40 & 0.10  &  1 & $f=5.42$\,d$^{-1}$    &     & 7,8 \\
$\alpha$\,Pyx   & 22900 &  300 & 3.60 & 0.05 &  2 & $\beta$\,Cep candidate  &      & 9 \\
$\epsilon$\,Lup & 19300 &  900 & 3.89 & 0.20 &  3 & $\beta$\,Cep candidate  &      & 9 \\
V2052\,Oph      & 23000 & 1000 & 4.00  & 0.20 &  1 & $f=7.148$\,d$^{-1}$   & 6.7  & 10 \\
V1449\,Aql      & 24500 & 1000 & 3.45 & 0.15 &  4 &  $f=5.48694$\,d$^{-1}$ & 44.0 & 4 \\
$\beta$\,Cep    & 25000 & 1000 & 3.80 &0.10  & 1   & $f=5.250$\,d$^{-1}$  & 21   & 11  \\
\hline
\end{tabular}
\tablefoot{
1 -- Lefever et al.\ (\cite{Lefever2010}),
2 -- Przybilla et al.\ (\cite{Przybilla2010}),
3 -- Hubrig et al.\ (\cite{Hubrig2009}),
4 -- Briquet et al.\ (\cite{Briquet2009}),
5 -- Aerts et al.\ (\cite{Aerts1992}),
6 -- Saesen et al.\ (\cite{Saesen2006}),
7 -- Heynderickx (\cite{Heynderickx1992}),
8 -- Heynderickx et al.\ (\cite{Heynderickx1994}),
9 -- Telting et al.\ (\cite{Telting2006}).
10 -- Neiner et al.\ (\cite{Neiner2003}),
and 11 -- Aerts et al.\ (\cite{Aerts1994})
}
\end{table*}

\begin{figure}
\centering
\includegraphics[width=0.40\textwidth]{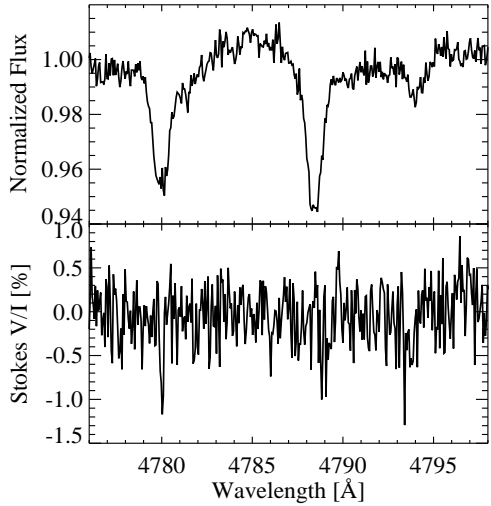}
\caption{
An example of Stokes~$I$ and Stokes~$V$ spectra observed at phase 0.648 in the wavelength 
region $\lambda\lambda$ 4775--4800
containing several \ion{N}{II} spectral lines.
}
\label{fig:new}
\end{figure}

\begin{figure}
\centering
\includegraphics[width=0.35\textwidth]{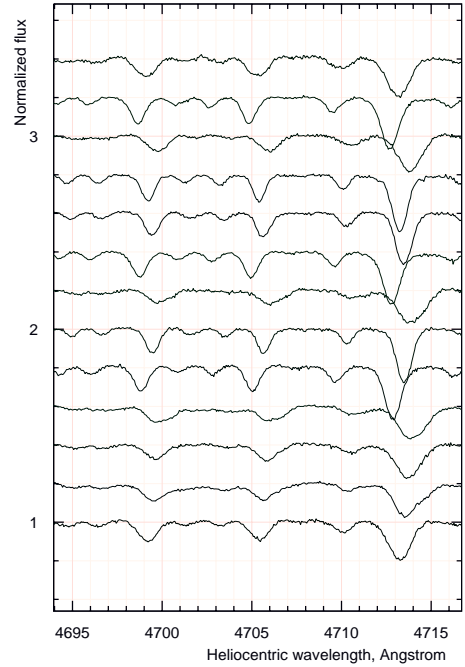}
\caption{
Line profile variability of \ion{O}{ii} $\lambda\lambda$4699, 4705, 4710,
and \ion{He}{i} $\lambda$4713
over the rotation cycle on SOFIN spectra. The spectra are presented with phases increasing from bottom to top
and offset in the vertical direction for clarity.
}
\label{fig:krivky}
\end{figure}


\begin{thebibliography}{}

\bibitem[1992]{Aerts1992}
Aerts, C., de Pauw, M., \& Waelkens, C.\ 1992,
A\&A, 266, 294

\bibitem[1994]{Aerts1994}
Aerts, C., Mathias, P., Gillet, D., \& Waelkens, C.\ 1994,
A\&A, 286, 109



\bibitem[2009]{Belkacem2009}
Belkacem, K., Samadi, R., Goupil, M.-J., et al.\ 2009,
Science, 324, 1540

\bibitem[2009]{Briquet2009}
Briquet, M., Uytterhoeven, K., Morel, T., et al.\ 2009,
A\&A, 506, 269

\bibitem[1989]{CroweGillet1989}
Crowe, R., \& Gillet, D.\ 1989,
A\&A, 211, 365

\bibitem[2009]{Degroote2009}
Degroote, P., Briquet, M., Catala, C., et al.\ 2009,
A\&A, 506, 111

\bibitem[2010]{Diaz2010}
D\'iaz, C.~G., Gonz\'alez, J.~F., Levato, H., \& Grosso M.\ 2010,
A\&A, {\sl accepted}, arXiv:1012.4858

\bibitem[2000]{Henrichs2000}
Henrichs, H.~F., Neiner, C., Hubert, A.~M., et al.\ 2000,
ASPC, 214, 372

\bibitem[1992]{Heynderickx1992}
Heynderickx, D.\ 1992,
A\&AS, 96, 207

\bibitem[1994]{Heynderickx1994}
Heynderickx, D., Waelkens, C., \& Smeyers, P.\ 1994,
A\&AS, 105, 447

\bibitem[2006]{Hubrig2006}
Hubrig, S., Briquet, M., Sch\"oller, M., et al.\ 2006,
MNRAS, 369, L61

\bibitem[2009]{Hubrig2009}
Hubrig, S., Briquet, M., De Cat, P., et al.\ 2009,
AN, 330, 317

\bibitem[2011]{Hubrig2011}
Hubrig, S., Ilyin, I., Sch\"oller, M., et al.\ 2011,
ApJ, 726, L5

\bibitem[2000]{Ilyin2000}
Ilyin, I.\ 2000,
Ph.D.\ Thesis, University of Oulu, Finland

\bibitem[2010]{Lefever2010}
Lefever, K., Puls, J., Morel, T., et al.\ 2010,
\aap, 515, A74

\bibitem[1991]{Mathias1991}
Mathias, P., Gillet, D., \& Crowe, R.\ 1991,
A\&A, 252, 245 

\bibitem[1994]{Mathys1994}
Mathys, G.\ 1994,
A\&AS, 108, 547

\bibitem[2007]{Miglio2007}
Miglio, A., Montalb\'an, J., \& Dupret, M.-A.\ 2007,
Comm.\ in Asteroseismology, 151, 48 

\bibitem[2006]{Morel2006}
Morel, T., Butler, K., Aerts, C., et al.\ 2006,
A\&A, 457, 651

\bibitem[2008]{Morel2008}
Morel, T., Hubrig, S., \& Briquet, M.\ 2008,
A\&A, 481, 453

\bibitem[2003]{Neiner2003}
Neiner, C., Henrichs, H.~F., Floquet, M., et al.\ 2003,
A\&A, 411, 565

\bibitem[1992]{Press92}
Press, W.~H., Teukolsky, S.~A., Vetterling, W.~T., \& Flannery, B.~P.\ 1992,
Numerical Recipes, 2nd ed.\ (Cambridge University Press: Cambridge)

\bibitem[2008]{Przybilla2008}
Przybilla, N., Nieva, M.-F., \& Butler, K.\ 2008,
ApJ, 688, L103

\bibitem[2010]{PrzybillaNieva2010}
Przybilla, N., \& Nieva, M.~F.\ 2010,
arXiv:1011.5977

\bibitem[2010]{Przybilla2010}
Przybilla, N., Firnstein, M., Nieva, M.~F., et al.\ 2010,
A\&A, 517, A38

\bibitem[2006]{Saesen2006}
Saesen, S., Briquet, M., \& Aerts, C.\ 2006,
Comm.\ in Asteroseismology, 147, 109

\bibitem[2008]{Scu2008}
Scuflaire, R., Th\'eado, S., Montalb\'an, J., et al.\ 2008,
Ap\&SS, 316, 83 

\bibitem[1977]{Seber77}
Seber, G.~A.~F.\ 1977,
Linear Regression Analysis (Wiley: New York) 

\bibitem[2006]{Telting2006}
Telting, J.~H., Schrijvers, C., Ilyin, I.~V., et al.\ 2006,
A\&A, 452, 945

\bibitem[1999]{Tuominen1999}
Tuominen, I., Ilyin, I., \& Petrov, P.\ 1999,
in Astrophysics with the NOT, eds.\ H.\ Karttunen \& V.\ Piirola, University of Turku, Tuorla Observatory, 47

\bibitem[1998]{Waelkens1998}
Waelkens, C., Aerts, C., Kestens, E., et al.\ 1998,
A\&A, 330, 215















%
%






%

\end{thebibliography}
\end{document}